\documentclass[5p,authoryear,twocolumn]{elsarticle}
\usepackage{graphicx,natbib}
\usepackage{amsmath,amssymb,latexsym}
\usepackage{multirow}
\usepackage{natbib}
\usepackage{color}
\definecolor{lila}{rgb}{0.5,0,1}
\definecolor{grau}{rgb}{0.5,0.5,0.5}

\newcommand{\Teff}{T_{\rm eff}}
\newcommand{\Teq}{T_{\rm eq}}
\newcommand{\Tint}{T_{\rm int}}
\newcommand{\Leff}{L_{\rm eff}}
\newcommand{\Leq}{L_{\rm eq}}
\newcommand{\Lint}{L_{\rm int}}

\newcommand{\Fint}{F_{\rm int}}
\newcommand{\ME}{M_{\oplus}}
\newcommand{\RE}{R_{\oplus}}

\newcommand{\MU}{M_{\rm U}}

\newcommand{\gccm}{$\,\rm g\, cm^{-3}$}
\newcommand{\Tcore}{T_{\rm core}}

\newcommand{\PLB}{P^{(LB)}}
\newcommand{\TLB}{T^{(LB)}}
\newcommand{\fig}{Fig.$\:$}

\journal{Icarus}

\begin{document}
\begin{frontmatter}
\title{Uranus evolution models with simple thermal boundary layers\\
\vspace{0.2cm}
\footnotesize\textit{Accepted to Icarus, 08 April 2016}}
 
\author[URos,UCSC]{N.~Nettelmann\corref{cora}}
\ead{nadine.nettelmann@uni-rostock.de}
\author[HS1,SIP]{K.~Wang}
\author[UCSC]{J.J.~Fortney}
%\ead{jfortney@ucolick.org}
\author[LLNL]{S.~Hamel}
%\ead{hamel2@llnl.gov}
\author[HS2,SIP,UCB]{S.~Yellamilli}
\author[URos]{M.~Bethkenhagen}
%\ead{mandy.bethkenhagen@uni-rostock.de}
\author[URos]{R.~Redmer}
%\ead{ronald.redmer@uni-rostock.de}
\address[URos]
{Institute for Physics, University of Rostock, 18051 Rostock, Germany}
\address[UCSC]
{Department of Astronomy and Astrophysics, University of California, Santa Cruz, CA 95064, USA}
\address[LLNL]
{LLNL, Livermore, CA, USA} 
\address[HS1]
{Castilleja High School, Palo Alto, CA, USA}
\address[HS2]
{Saratoga High School, Saratoga, CA, USA}
\address[UCB]
{University of California, Berkeley, CA, USA}
\address[SIP]
{Science Internship Program, UCSC 2014}
\cortext[cora]{Corresponding Author} 

\begin{abstract}
The strikingly low luminosity of Uranus ($\Teff\simeq\Teq$) constitutes a long-standing challenge to our understanding of 
Ice Giant planets. Here we present the first Uranus structure and evolution models that are constructed to agree 
with both the observed low luminosity and the gravity field data.   
Our models make use of modern \emph{ab initio} equations of state at high pressures for the icy components water, 
methane, and ammonia.
Proceeding step by step, we confirm that adiabatic models yield cooling times that are too long, even when uncertainties 
in the ice:rock ratio (I:R) are taken into account. 
We then argue that the transition between the ice/rock-rich interior and the H/He-rich outer envelope should be stably stratified. Therefore, we introduce a simple thermal boundary and adjust it to reproduce the low luminosity. Due to this thermal boundary, the deep interior of the Uranus models are up to 2--3 warmer than adiabatic models, necessitating the presence of rocks in the deep interior with a possible I:R of $1\times$ solar.
Finally, we allow for an equilibrium evolution  ($\Teff\simeq\Teq$) that begun prior to the present day,
which would therefore no longer require the current era to be a ''special time'' in Uranus' evolution.
In this scenario, the thermal boundary leads to more rapid cooling of the outer envelope. When $\Teff\simeq\Teq$ is reached, 
a shallow, subadiabatic zone in the atmosphere begins to develop. Its depth is adjusted to meet the luminosity constraint. This work provides a simple foundation for future Ice Giant structure and evolution models, that can be improved by properly treating the heat and particle fluxes in the diffusive zones.
\end{abstract} 

\end{frontmatter}
%\linenumbers

%%%%%%%%%%%%%%%%%%%%%%
\section{Introduction}

Uranus is an interesting planet in many aspects.  Uranus closely resembles its neighbor Neptune in many physical properties such as mass, size, rotation rate, surface temperature, effective temperature, magnetic field geometry, and atmospheric zonal flow pattern, suggesting both planets represent a common class of planets.
Moreover, its 21 year seasons provide unique conditions for investigating irradiation driven atmospheric dynamics and 
horizontal energy transfer \citep{Ingersoll87,Allison91}. 
Furthermore, Uranus' atmospheric thermal emission is in or close to equilibrium with the solar incident flux \citep{Pearl90,Conrath91} 
---similar to the Earth atmosphere, although the interior of Uranus is thought to consist primarily of conducting, fluid ices  
\citep{Podolak91,Hubbard95} rather than of solid, poorly conducting rocks like terrestrial planets.  
Finally, with its mass of 14.5 $\ME$ and mean density of 1.3\gccm, Uranus is our nearest analog to the most frequent currently detectable exoplanets, which are the low-mass low-density exoplanets of radius 2--4 $\RE$ on tight ($\lesssim 100$ days) orbits \citep{Fressin13, Fortney13, Batalha14}, also labeled mini-Neptunes or warm Neptunes depending on their mass. 
Their bulk composition is usually derived from adiabatic, quasi-homogeneous interior structure and evolution models \citep{RogSea10,Lopez12}. However, there are strong indications that  these assumptions do not hold for Uranus, which is the topic of this paper.
 
Adiabatic, quasi-homogeneous, and water-rich structure and evolution models can well explain several observed properties of Uranus and Neptune, for instance their gravity field \citep{Podolak95,Hubbard95,Helled11}. However, the assumption of adiabaticity and quasi-homogeneity appears to fail in some respects. In particular, such models cannot explain the difference in intrinsic luminosities by a factor of $\sim 10$ between Uranus and Neptune \citep{Guillot05}. Moreover, no  current model is consistent with Uranus' faintness, while the same model assumptions work well for Neptune \citep{FN10,Fortney11,N13UN}. Furthermore, the measured D/H ratio in Uranus' troposphere is lower than would be expected if the deep ice-rich interior and atmosphere were fully mixed, and if deuterium entered the planet primarily through cometary ices \citep{Feuchtgruber13}. These shortfalls can be considered strong indication that the assumptions of homogeneity and adiabaticity do not hold in the entire interior of the Ice Giant Uranus.

Of course, the interior of Uranus and Neptune may be more complex than has been accounted for in planet models so far. For instance, a property that has been suggested to overcome the shortfalls is the occurrence of inhomogeneous zones that act as thermal boundary layers (TBL) \citep{MarGom95,Hubbard95}, perhaps as a result of the formation process \citep{Stevenson85,PodoHell12}. Inhomogeneous zones can strongly influence the thermal evolution, the present luminosity, the entire temperature profile, and thus the inferred bulk composition.

In this paper, we consider the evolution of Uranus under the assumption that one inhomogeneous zone exists that acts as a
thermal barrier to the deep internal heat flow. We represent it in a very simplified manner by a jump in temperature and  show that this property can indeed easily explain the low luminosity. In addition, we suggest the presence of a shallow radiative zone in Uranus' deeper atmosphere as a result of  its equilibrium evolution with the solar incident flux.

We arrive at our conclusions by considering three classes of thermal evolution models.
They all correspond to a ''hot-start'' after planet formation, where the planet cools down from an 
initial state of high internal specific entropy \citep{Marley07}.
After describing the input data in Section \ref{sec:methods} we show in Section \ref{sec:evol_ad} that taking 
into account the uncertainty in ice-to-rock ratio (I:R) in conventional, adiabatic models (class I)  does not 
solve the faintness problem.

In Section \ref{sec:SIP} we argue that common Uranus structure models that meet the gravitational field data actually yield evidence for a stably stratified boundary at the transition between the H/He rich outer and the ice-rock rich inner envelope at $\sim 0.1$ Mbar \citep{Podolak91,N13UN}. While its width is unconstrained by the gravity data \citep{MarGom95,Podolak95,Helled11}, we here assume a thin superadiabatic zone. A thin zone can be motivated by the observed multipolar magnetic field and corresponding dynamo models, which require the presence of convective motions in a conducting fluid  in the vicinity of the transition zone \citep{StanBlox06,Soderlund13}. 
In our class II evolution model we simply adjust the temperature jump across the TBL to match the observed luminosity.
Class I and II evolution models rely on the assumption that Uranus is now beginning to evolve in equilibrium with the solar incident flux.

In Section \ref{sec:evol3} we consider the --we think-- more likely scenario that Uranus's atmosphere has been in a state of equilibrium evolution for already some time. As a consequence, a shallow radiative, subadiabatic zone develops in the deeper atmosphere and exists today in addition to the  thermal boundary at $\sim 0.1$ Mbar (class III).
Finally, in Section \ref{sec:structure} we investigate the effect of class II and III evolution models to the composition and the possible thermodynamic phases of water, ammonia, and methane.
%%% 5,6,7 %%%
Section \ref{sec:summary} outlines directions for further improvements.

%%%%%%%%%%%%%%%%%%%%%%%%%%%%%%%%%%%%%%%%%%%
\section{Input data}\label{sec:methods}

%%%%%%%%%%%%%%%%%%%%%%%%%%%
\subsection{Energy balance}

The energy balance equation for the observable total luminosity $L$, the incident luminosity $\Leq$, and the intrinsic luminosity $\Lint$ simply reads $L-\Leq=\Lint$. By Stefan-Boltzmann's law, we have $L=4\pi R_p^2\,\sigma\Teff^4$ and
$\Leq=4\pi R_p^2\,\sigma \Teq^4$, while $\Lint=-\int_0^{M_U}dm\, T(ds/dt)$. 
The equilibrium temperature $\Teq$ can be written as $\sigma\Teq^4 = f(1-A) F_{\star}(t,a)$, where $F_{\star}(t,a)=\sigma T_{\star}^4(t) \times(R_{\star}^2(t)/a^2)$ is the solar flux at orbital distance $a$ and time $t$, and $f$ a geometry factor, usually taken to be 1/4 for a planet where the day-side (Uranus: the summer-side) irradiation is distributed over the entire surface. For the rapid rotator Neptune, the energy from solar incident flux gets easily distributed to the night-side by zonal winds, while for the oblique Uranus meridional wind are argued to do that job \citep{Ingersoll87}.  From analysis of disk brightness and thermal emission measurements during the Voyager 2 flyby, the effective temperature $\Teff=59.1\pm 0.3$\ K and the Bond Albedo $A=0.300\pm 0.049$ could be determined \citep{Pearl90}. For $a=19.2$\ AU \citep{Arridge11}, and a Solar constant $S=0.1361\rm W/cm^2$ the latter quantity yields $\Teq = 58.1\pm 1.0$\ K for Uranus today, and therefore $L/\Leq = 1.07^{+0.1}_{-0.09} \simeq 1$, or $\Fint=45\pm 47$ erg/s/cm$^2$, unlike for any other giant planet in the solar system. This implies that the intrinsic energy loss of Uranus is consistent with being zero, but perhaps it is slightly higher than the earlier estimate of \citet{Pearl90} who used a higher Solar constant value of that time. 
In our thermal evolution calculations, we use $A=0.300$ and fit $\Teff$ to within its $1\sigma$ error bars.

%%%%%%%%%%%%%%%%%%%%%%%%%%%%%%%%%%
\subsection{Equations of state}\label{sec:EOS}

To model the internal structure and thermal evolution of Uranus we apply equations of state (EOS) for 
H, He, H$_2$O, CH$_4$, NH$_3$, basalt, and rocks. For hydrogen, helium, and water we use the EOSs described in 
\citet{Nett08Jup}, which are, respectively, H-REOS.1, He-REOS.1, and H2O-REOS.1.
For rocks in the inner envelope ($P\gtrsim 0.1$~Mbar, $T\gtrsim 2000$~ K) we use the SESAME EOS table 7350 \citep{SESAME} for basalt, where ''basalt'' encompasses a variety of heavy elements. The table contains data for $T\geq 1160$ K.

\begin{figure}
\centering
\includegraphics[width=0.45\textwidth]{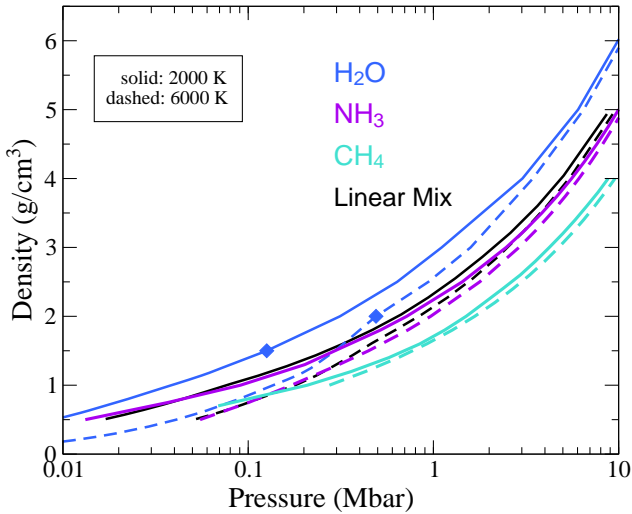}
\caption{\label{fig:isoTs_ices}Isotherms for 2000 K (solid) and 6000 K (dashed) of the ab initio EOS for NH$_3$ (violet), CH$_4$ (cyan), and H$_2$O (blue). For water, ab initio data are used for pressures higher or equal to the points marked by \emph{diamonds}. The \emph{black} curves are for the linear mixing approximation and mass mixing ratios 
$\rm H_2O:CH_4:NH_3 \approx 7.7\,:\,4\,:\,1$. }
\end{figure}

For light ices CH$_4$ and NH$_3$ we apply new ab initio EOS data. 
%%% NH3 %%%
The ab initio EOS of ammonia is presented in \citet{Bethkenhagen13}. We have 
extended that table toward a rectangular grid in $T$--$\rho$ space, which reaches from 1,000 to 10,000\,K 
in steps of 1000\,K (9000\,K omitted) and from 0.5 to 5.25\,\gccm. This corresponds to
$0.01 \lesssim P\mbox{(Mbar)} \lesssim 10$.
%%% CH4 %%%
The methane EOS (Bethkenhagen et al., in prep.) currently spans the range of $2000 \leq T\mbox{(K)} \leq 10,000$ and $0.05 \lesssim P\mbox{(Mbar)} 
\lesssim 10$ along the Uranus adiabat. We use these methane data for 2000 and 2750\,K within 0.7--4\,\gccm, for 3000--7000\,K in steps of 1000\,K within 1--4\,\gccm, and for 10000\,K within 1.2--2.8\,\gccm.
Among these EOSs for the light ices, methane has the stiffest $P$--$\rho$ relation and water the softest one, as can be 
seen from the isotherms shown in \fig\ref{fig:isoTs_ices}. At relevant pressures of 1--10\,Mbar in the planetary 
inner envelope, the density of CH$_4$ is $\sim 0.5$--{1\,\gccm} lower than that of NH$_3$, and the same increase is 
seen between NH$_3$ and H$_2$O. We thus expect a mixed CH$_4$-NH$_3$-H$_2$O EOS of solar composition to be 
less dense than a pure water EOS. Indeed, as Figure \ref{fig:isoTs_ices} shows, the resulting 
density of a mixture of ices of mass mixing ratios 
($\rm  Z_{CH_4} : Z_{NH_3} : Z_{H_2O}$ $\approx 4\,:\,1\,:\,7.7$) 
is closer to that of ammonia than to that of water.

%%%%%%%%%%%%%%
%%%%%%%%%%%%

%%%%%%%%%%%%%%%%%%%%%%%%%%%%%%%%%%%%%%%%%%%
%\hspace*{1cm}\\
%{\bf \large[Figure \ref{fig:isoTs_ices}]}
%\hspace*{1cm}\\
%%%%%%%%%%%%%%%%%%%%%%%%%%%%%%%%%%%%%%%%%%

The above mentioned ranges cover the entire inner envelope of present Uranus if it is adiabatic. For potentially higher deep internal temperatures of superadiabatic models and for its early evolution, we use extrapolated 20,000 K isotherms for the EOSs of CH$_4$ and NH$_3$.

As we have not constructed smooth interpolations to possibly available EOS tables for methane, ammonia, and rocks at lower pressures and temperatures, we apply the light ice and basalt EOSs in only the inner envelope of Uranus, where $T\geq 2000$ K. As in our previous Uranus models, ices in the outer envelope  are represented by the water EOS H2O-REOS.1.
Finally, rocks in the core are represented by the rock EOS of \citet{HM89}, which is supposed to describe a bulk-Earth-like mixture of 38\% SiO$_2$, 25\% MgO, 25\% FeS, 12\% FeO under Jovian core conditions. The composition of rocks in the real planet is unknown. Definitions of our layer labels used throughout this paper are given in Table \ref{tab:layerdefs}.

%%%%%%%%%%%%%%%%%%%%%%%%%%%%%%%
%{\bf [Table \ref{tab:layerdefs}]}

%%% definition of the layer labels used throughout the paper %%%
\begin{table}
\setlength{\tabcolsep}{2pt}
\begin{footnotesize}
\centering
\caption{\label{tab:layerdefs}Layer definitions of our Uranus models.}
\rotatebox{0}{
\begin{tabular}{cccc}
\hline
\hline
No. & Label &  Pressure range & property\\
\hline
1 & atmosphere & 1--1000 bars & H/He-rich weather layer\\
1 & outer env. & 1 bar--$P^{(LB)}$  & H/He-rich, convec., adiab.\\
TBL  & therm.~boundary & around $P^{(LB)}$ &  radiative \\ 
2 & inner env. & $P^{(LB)}$--$P_{\rm core}$  & ice-rich, convec., adiab.\\
3 & core & $P_{\rm core}$--$P_{\rm center}$  & rocks\\
2+3 & deep interior & $P^{(LB)}$--$P_{\rm center}$ & ice/rock-rich\\
\hline
\\
\multicolumn{4}{c}{ 
%\begin{minipage}{0.9\textwidth}
%\vspace{2mm}
$P^{(LB)}$ $\approx$0.1 Mbar, % (pressure at the transition between inner and outer envelope), 
$P_{\rm core}$ $\approx$5.5 Mbar,
$P_{\rm center}$ $\approx$ 9 Mbar.}
%\end{minipage}
\end{tabular}}
\end{footnotesize}
\end{table}

%%%%%%%%%%%%%%%%%%%%%%%%%%%%%%%

%%%%%%%%%%%%%%%%%%%%%%%%%%%%%%%%%%%%%%
\subsection{Baseline structure models}\label{sec:struc_ad}

We apply two structure models for our thermal evolution computations.

For our adiabatic evolution computations in Section \ref{sec:evol_ad} we use a standard structure model
as presented in \cite{N13UN}. It has three layers consisting of a central rock core, and two adiabatic and homogeneous
envelopes. The outer envelope (indexed by (1)), is less enriched in heavy elements than the inner one (indexed by (2)). Heavy elements in the inner envelope are represented either by water or by basalt, in the latter case leading to a maximum rock mass fraction there of $Z^{(2)}_{\rm rocks}\approx 0.75$ in agreement with \citet{Helled11}.

For our evolution calculations with a TBL in Sections \ref{sec:SIP} and \ref{sec:evol3} we assume similar  three-layer structures. Application of the EOSs for the light ices ammonia and methane allows for, but not requires, a H/He free deep interior. For our potentially very warm class II and III evolution model we assume no H/He in the inner envelope, solar proportions O:C:N, and use the I:R ratio to adjust the planet mean radius of the spherical models.

In the real planet, the O:C:N ratio could be different from solar. A sub-solar C:O can arise if the planet accreted solid material from orbital distances between the snow lines of water and methane \citep{Oberg11}. On the other hand, the atmosphere of Uranus appears to be enriched in C/O as the observed values of C/H is about $80\times$ solar \citep{GuiGau14}, while, according to our structure models, the O/H ratio needed to fit the gravitational harmonic $J_4$ is less than $30\times$ solar. However, we do not consider $\rm C/O >1$ a good assumption for the bulk planet. As we will see below, varying the C/O ratio from 0 to $1\times$ solar reduces the cooling time by $\sim 0.5$ Gyrs, so that we estimate the entire uncertainty in cooling time due to an uncertainty $\rm 0<C/O<1$ to be $\simeq 1$ Gyr.

%%%%%%%%%%%%%%%%%%%%%%%%%%%%%%%%%%%%%%%%%
\subsection{Model atmosphere}

Model atmospheres relate internal structure to luminosity during the thermal evolution.  Here we apply the description of \citep{Guillot95}  to the \citet{Graboske75} model atmosphere, noting that a more modern model atmosphere grid exists \citep{Fortney11}. The latter one can be considered more accurate as it is more densely spaced in planet surface gravities and computed for high-Z atmospheres appropriate for Uranus. However, it does not reproduce the observationally derived 1-bar surface temperatures. Nevertheless, the grids have been shown to yield very similar cooling times for Neptune  and to confirm the issue of a too long cooling time for Uranus \citep{Fortney11}. Since in this paper we are interested in general concepts for more self-consistent structure and evolution models that explain both the gravity data and the low luminosity of Uranus, we think that application of the former grid is sufficient at this stage. 
We use it with a scaling factor between effective and 1-bar temperature of $K=1.48$, which reproduces Uranus' measured effective temperature to within its $1\sigma$ limit. 

%%%%%%%%%%%%%%%%%%%%%%%%%%%%%%%%%%%%%%%%%
\subsection{Uncertainties in input data}

Beside uncertainties in the available model atmospheres, the current observational uncertainties in $\Teff$ and $A$, as well as the EOS of the light ices, may influence the computed cooling time of Uranus. For instance, \citet{Fortney11} found a shortening in cooling time by 2 Gyrs when using the $1\sigma$ upper limit in intrinsic heat flux for Uranus, corresponding to the $1\sigma$ upper limit in $A$ or the $4\sigma$ upper limit in $\Teff$, and \citet{FN10,Fortney11} found a shortening of Neptune's computed cooling time down to $\tau_{\rm N} \simeq \tau_{\odot}$ by using advanced EOS for water compared to previous ice EOS in use \citep{HMacF80}.

Detailed investigations of the influences of the EOS of the light ices methane and ammonia, of the uncertainties in $\Teff$ and $A$, which together determine the possible intrinsic heat loss, as well as of the uncertainty in the model atmosphere will be subject to future work.

%%%%%%%%%%%%%%%%%%%%%%%%%%%%%%%%%%%%%%%%%%%%%%%%%%%%%%%%%%%%%%%%%%%%
\section{Evolution I: adiabatic models}\label{sec:evol_ad}

%%%%%%%%%%%%%%%%%%%
\subsection{Review}

%%% some History: Hubbard 1978 %%%
\citet{Hubbard78} was the first to compute thermal evolution models for Uranus and Neptune. Assuming both planets evolved homogeneously over time, in his notation \emph{convective-cooling}, as such an assumption worked successfully in explaining Jupiter's observed luminosity, he applied a scaled Jupiter model, in which planet mass, radius, central density, and heat capacity are adjusted to the ice giants values, and the computed cooling would mainly depend on the planet's $\Teff$ and $\Teq$. With the uncertainty in the planets' $\Teff$ of $\pm 2$~K at that time, the computed cooling times of both Uranus and Neptune were in agreement with the age of the solar system. \citet{Hubbard78} argued that Uranus' observed faintness was caused by the strong insulation, $\Teff^{(U)}/\Teq^{(U)}\gtrsim 1$, which would eventually lead, perhaps already has led, to the breakdown of the internal thermally-driven convection. In contrast,  $\Teff^{(N)} > \Teq^{(N)}$ for Neptune because of its larger orbital distance. 

%%% some History: Hubbard MacFarlane 1980 %%%
\citet{HMacF80} applied contemporary physical EOS for ices and rocks and assumed a three-layer structure with a solar-composition  envelope, an ice shell, and a rock core with solar bulk I:R ratio to model the structure and evolution of Uranus  and Neptune. Although taking some uncertainty in the specific heat of ices into account, they found too long cooling times for both planets. Only for an ice-free, rock-rich interior could the low luminosity of Uranus be matched.
On the other hand, while some ice-depletion might be consistent with cosmochemical properties of giant planet formation \citep{Stevenson85}, an entirely ice-free interior is considered not consistent with the observed atmospheric CH$_4$ enrichment \citep{HMacF80} unless it does not originate from the deep interior.

%%%%%%%%%%%%%%%%%%%%%%%%%%%%%%%%%%%%%%%%%%%%%%%%%%%%%%%%%%%%%%%%%%%%%%%%%%%%%%%%%%%%%%%
\subsection{Results for rock-rich interior}\label{sec:evol_rocks}

%%%%%%%%%%%%%%
\begin{figure}
\centering
\includegraphics[width=0.45\textwidth]{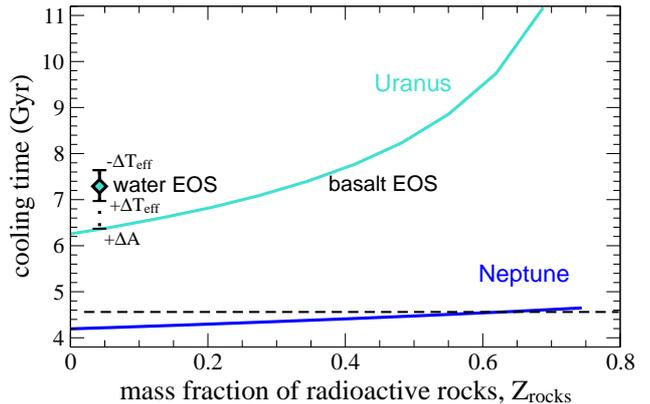}
\caption{\label{fig:evol_Zrocks}(Color online) Cooling time of Uranus (\emph{cyan}) and Neptune (\emph{blue}), as a function of the mass fraction of radioactive heavy elements in the deep envelope (\emph{x-axis}), which are represented by a basalt EOS. \emph{Diamond}: using water as proxy for heavy elements; its error bars: uncertainty due to the $1\sigma$ limits in $\Teff$ and albedo. 
This figure shows the strong influence of the energy of radioactive rocks on the cooling time of conventional, adiabatic Uranus evolution models (class I).}
\end{figure}
%{\bf \large[Figure \ref{fig:evol_Zrocks}]}
%\hspace*{1cm}\\

Since these earlier works the available EOSs of ices and rocks have changed. Application of the ab initio water EOS H2O-REOS.1 and of the SESAME water EOS to icy models of Neptune now reach the measured luminosity at the proper cooling time \citep{FN10,Fortney11}.
Assuming a standard adiabatic three-layer structure as well, we here re-investigate the effect of I:R ratio on the cooling time of Uranus by varying the mass fraction of rocks in the inner envelope, $Z_{rocks}^{(2)}$. Unlike \citet{HMacF80} we include the energy from the decay of radioactive elements associated with the rock-component, for which we assume Earth-like abundances at present time and, according to their decay rates, higher abundances at earlier times (\citealp[see][for details]{NFKR11}).
Figure \ref{fig:evol_Zrocks} shows the result.

Apparently, the effect of the additional luminosity from the radioactive elements is stronger than the considered uncertainty in specific heat due to the uncertainty in the I:R ratio. Therefore we conclude that the uncertainty in the I:R ratio does not provide a solution to the low luminosity of Uranus.
In contrast, we find that the computed cooling time of Neptune stays close to the solar system age, $\tau_{\odot}$, whatever the rock mass fraction.

%%%%%%%%%%%%%%%%%%%%%%%%%%%%%%%%%%%%%%%%%%%%%%%%%%%%%%%%%%%%%%%%%%
\section{Evolution II: adjusted thermal boundary layer}\label{sec:SIP}

For Uranus, all published structure models that meet the measured $J_2$ and $J_4$ values require the heavy element abundance to increase with depth, where both a continuous \citep{MarGom95,Podolak00,Helled11,Fortney13} or a discontinuous \citep{Hubbard95, N13UN} increase is allowed by the gravity data. In other words, a fully homogeneous interior would be inconsistent with the measured gravity data.  
Therefore, we assume that a TBL exists at the location of the layer boundary of our adiabatic, layered  structure models at $\approx 0.1$ Mbar.

%%%%%%%%%%%%%%%%%%%%%%%%%%%%%%%%
\subsection{Origin of the layer boundary}

The origin of a compositional difference could be related to the formation process \citep{PodoHell12}. With $\approx 13\ME$, the planet mass beneath the layer boundary is in agreement with predictions of critical core masses for run-away gas accretion \citep{Pollack96,Mordasini12}. In this scenario, the layer boundary of our structure models, which separates the high-Z deep interior from the H/He-rich outer part, might correspond to the core-mantle boundary of conventional giant planet models, which are thought to have accumulated about 5--15$\ME$ of high-Z material \citep{Pollack96}.
In another scenario, the planet may have formed inhomogeneously, i.e.~with compositional differences being characteristic rather than the exception, and then its outer part become homogenized through giant impacts \citep{Stevenson86}. Upon giant impact, the deposited energy can lead to homogenization of the outer shells, while the deposited angular momentum can cause an axial tilt. Both processes depend on the impact parameter of the infalling planetesimals \citep{PodoHell12}. Therefore, \citet{Stevenson86} suggested an oblique impact at a late stage of Uranus' formation to explain its high axial tilt of 98 degrees, though it may have required more than one shot \citep{Morbi12}. Our structure models are consistent with that scenario in that they have a rather small homogeneous outer envelope (perhaps due to a late grazing impact) ontop a compositionally different region (perhaps a remnant of inhomogeneous formation). On the other hand, we do not assume a fully inhomogeneous deep interior as proposed in the latter formation scenario. One could, however, bring our model for the deep interior of Uranus easily into agreement with the formation scenario of \citet{Stevenson86} by assuming  radial giant impacts onto proto-Uranus \emph{before} accretion was completed.

An initial compositional gradient can then stabilize itself through the suppression of convection. While in the absence of overturning convection, particle exchange might still occur through semiconvection or diffusion, their efficiencies in distributing particles (and heat) are generally thought to be much smaller than that of convection \citep{Stevenson85,LissStev07}, in particular when the compositional gradient in the giant planet is very steep \citep{Vazan15}. Here we make the simplifying assumption that no particles are transported across the layer boundary so that the compositional difference between the two envelopes stays constant over time.

Still, the layer boundary of our structure models may have started off with a gradient from the formation process, and today be of finite width rather than sharp as assumed here. For simplicity, we take the compositional difference between the layers given but allow for a small, variable width $\Delta P$ of it. For the respective mean molecular weights $\mu$ in the outer envelope of composition $Z_{\rm H2O}^{(1)}$=0.08 we find $\mu^{(1)}=2.5$ g\ mol$^{-1}$, and for the inner envelope of composition $Z_{\rm H2O}^{(2)}=0.62$, $Z_{\rm CH4}^{(2)}=0.30$, $Z_{\rm NH3}^{(2)}=0.08$ we find $\mu^{(2)}=15.6$ g\ mol$^{-1}$. Next we use this information to argue that the layer is stable to convection. This property serves us to justify the assumption of a thermal boundary layer.

%%%%%%%%%%%%%%%%%%%%%%%%%%%%%%%%%%%%%%%%%%%%%%%%%%%%%%%%%%%%%%%%%%%%%%
\subsection{Stability of the thermal boundary layer}\label{sec:stabBL}

Given  the existence and compositional difference between the outer and the inner envelope, we here estimate the separating layer boundary's stability to convection. For that purpose we apply the Ledoux stability criterion in its 
general form \citep{KippWeig}, 
\begin{equation}\label{eq:ledoux}
\nabla_T < \nabla_{ad} + (\alpha_{\mu}/\alpha_{T})\nabla_{\mu} \:,
\end{equation}
which states that a system of particles is stable to convection as long as the change of density with height 
felt by an adiabatically expanding blob that got somehow displaced from its equilibrium position 
is \emph{smaller} than the change of density with height in the ambient fluid. With the help of the partial derivatives 
$\alpha_T^{(i)}=-\partial\log\rho /\partial \log T$  and $\alpha_{\mu}=\partial \log\rho/\partial \log\mu$, 
this criterion can be expressed in terms of the thermal gradient ($\nabla_T$) and the compositional gradient 
$\nabla_{\mu}=d\log\mu/d\log P$ as written in Equation \ref{eq:ledoux}.

We use this criterion to estimate the minimum change of temperature across the layer boundary, $\Delta T$ that would be necessary for convective instability, and then ask whether the resulting number adopts a realistic value. We approximate $\Delta T$ in dependence of the unknown width $\Delta P$ of the layer boundary by
\begin{equation}\label{eq:TLB}
\frac{ \Delta T }{ \Delta P } = \nabla_T\,\frac{ \TLB }{ \PLB }
\end{equation}
with $\TLB=2140$ K and $\PLB=0.11$ Mbar being the values at the layer boundary according to our icy adiabatic baseline structure model, see Section \ref{sec:struc_ad}. Since intermediate values within the boundary layer are unknown, we compute $\nabla_T$ in Eq.~(\ref{eq:TLB}) separately at the outer and at the inner edge with the help of Eq.~(\ref{eq:ledoux}),
\begin{equation}\label{eq:gradT}
\nabla_T^{(i)}=\bar{\nabla}_{ad} + \left(\alpha_{\mu}^{(i)} / \alpha_T^{(i)}\right) \nabla_{\mu}\:,
\end{equation}
where $i=1$ denotes the outer edge at $P^{(1)}=\PLB-0.5\Delta P$ and $i=2$ the inner edge at $P^{(2)}=\PLB+0.5\Delta P$. In Eq.~(\ref{eq:gradT}), the mean adiabatic gradient $\bar{\nabla}_{ad}$ is computed using
\begin{equation}\label{eq:gradad}
\frac{ \Delta T_{ad} }{ \Delta P } = \bar{\nabla}_{ad}\,\frac{ \TLB }{ \PLB } \:,
\end{equation}
where $\Delta T_{ad}$ is given by our adiabatic baseline structure model. The partial derivatives $\alpha_T^{(i)}$  
are taken at ($\PLB$, $\TLB$) and computed directly from the EOS tables for the different compositions of 
the two layers, yielding $\alpha_T^{(1)}=0.14$ and $\alpha_T^{(2)}=0.25$, while $\alpha_{\mu}$ is computed using 
\begin{equation}\label{eq:amu}
	\alpha_{\mu}^{(i)} = \frac{\rho^{(i)} (\PLB, \TLB)}{\mu^{(i)}}
	\sum_{j=1}^{N_i} \frac{\rho_j^{-1} - \rho_{\rm H}^{-1}}{\mu_j^{-1}-\mu_{\rm H}^{-1}} \:.
\end{equation}
In this context, $N_i$ denotes the number of components except hydrogen in layer numbers $i$, while $j$ enumerates the components.
We find $\alpha_{\mu}^{(1)}=2.2$ and $\alpha_{\mu}^{(2)}=2.0$. 
In Table \ref{tab:SIP_Ledoux} we give the parameter values as a function of $\Delta P$. In particular, we find $\bar{\nabla}_{ad}=0.425$ for an approximate mean adiabatic gradient 
across the layer boundary.

%%% TABLE %%%%%%%%%%%%%%%%%%%%%%%%
%\hspace*{1cm}\\
%{\bf\ [Table \ref{tab:SIP_Ledoux}]}
%\hspace*{1cm}\\

%%% values were computed by the SIP high school students, Kate and Shiv in Summer 2014 %%%
\begin{table}
\setlength{\tabcolsep}{4pt}
\caption{\label{tab:SIP_Ledoux}Values of the parameters used in Eqs.~\ref{eq:TLB}--\ref{eq:gradad}. }
\begin{tabular}{ccccccc}
\hline\hline
$\Delta P^{(LB)}$ & $\Delta T_{ad}$ & $\nabla_{\mu}$ & $\nabla_{T}^{(1)}$ & $\nabla_{T}^{(2)}$ 
& $\Delta T^{(LB,1)}$ & $\Delta T^{(LB,2)}$\\
(GPa) & (K) &  &  & & ($10^3$ K) & ($10^3$ K)\\ \hline
2 & 134 & 8.0 & 126 & 64.1 & 49.1 & 25.0 \\
4 & 290 & 4.0 & 63.3 & 32.3 & 49.4 & 25.2 \\
6 & 460 & 2.66 & 42.2 & 21.6 & 49.4 & 25.3\\
8 & 642 & 2.0 & 31.8 & 16.3 & 49.6 & 25.4 \\
10 & 858 & 1.6 & 25.5 & 13.2 & 49.7 & 25.7\\
\hline
\end{tabular}
\centering
\begin{minipage}{0.45 \textwidth}
\small The main results are the values of $\Delta T^{(LB,1)}$ and $\Delta T^{(LB,2)}$. These are our estimates for the change of temperature across the layer boundary using input values primarily from the outer edge of the layer boundary (index 1) or from its inner edge (index 2). Additional values used are: $P^{(LB)}=11$ GPa, $T^{(LB)}=2145$ K, $\Delta \mu=13.126$ g.
\end{minipage}
\end{table}

%%%%%%%%%%%%%%%%%%%%%%%%%%%%%%%%%%

Finally, $\Delta T$ can be obtained as a function of the assumed width $\Delta P$, as listed in the two rightmost columns of Table
\ref{tab:SIP_Ledoux}. We find $\Delta T = 25$--$50\times 10^3$~K.
The obtained values are nearly independent of the assumed uncertainty in layer boundary width for the widths considered. This is because  $\Delta P$ cancels out in Eq.~\ref{eq:gradT}, its influence on the estimated $\bar{\nabla}_{ad}$ is small, and the remaining coefficients are taken at same $(\PLB,\TLB)$.
Furthermore, the different compositions in the layers lead to only slightly different values of $\nabla_{ad}$, $\alpha_T$, and $\alpha_{\mu}$ so that the estimated values for $\Delta T$ from, respectively, the outer edge inward and the inner edge outward differ by a factor of 2 only.
We do not think that application of better approximations than the ones we made would change our result of 
$\Delta T \gtrsim$ (2--5)$\times 10^4$~K. This is a high number. We compare it to the theoretical amount of thermal energy in the planet. According to the Virial theorem, the mean thermal energy of a gravitationally bound, spherical body adopts a fraction $1/\xi \lesssim 1$ of the gravitational binding energy. This fraction is 1/2 for a monoatomic perfect gas or a degenerate nonrelativistic gas, and $1/3.2$ for a diatomic perfect gas \citep{Guillot05}. The gravitational binding energy of  the $13.5\ME$ ice-rock deep interior of Uranus, compressed to a radius of $R_{deep}\approx 3.2\RE$, is $E_{grav} = (3/5)\times$ $(GM_{deep}^2/R_{deep})$ 
$\approx 1.3\times 10^{34}\:$J. If converted to thermal energy, the increase in temperature of the initially cold ($\approx 50\,$K) gas cloud from which the planet's deep interior once formed could then at most be 
$E_{th} = M_{deep}\, c_v$  ($T_{deep} - 50\,$K)  $\approx (1/\xi) E_{grav}$. 
With specific heat $c_v=4$ J/gK, we estimate the maximum temperature to be 
$T_{deep} \lesssim 50\,{\rm K}$ $+ \frac{1}{\xi}\frac{13\times 10^{33}\rm J}{3.2\times 10^{29} {\rm J/K}}$ $\approx$ 
(1.3--2) $\times 10^{4}\,$K. \citet{Guillot05} furthermore points out that a significant fraction $(1-1/\xi)$ of the potential energy is radiated away during contraction; a better approximation is therefore 
$T_{deep}\lesssim \xi^{-2}\times 4\times 10^{4}$~K $\approx$ (0.4--1) $\times 10^{4}\,$K.
For comparison, \citet{Mordasini12} model the combined formation and thermal evolution of a 1 M$_{\rm Jup}$ mass planet at 5 AU and find a maximum central temperature of $7\times 10^4$~ K. Less-massive planets and planets at larger orbital distance such as Uranus are expected to reach lower maximum temperatures because of their lower total energy reservoir.

Because the estimated possible maximum value $T_{deep}$ is a little lower than our above estimate for the change of temperature $\Delta T\gtrsim 2\times 10^4$~K necessary for thermally driven convection across the layer boundary,  we suggest that the latter is an unrealistically high value. Therefore, we go on to suggest that convection does not operate across the layer boundary permanently, neither now nor in the past, and that therefore a TBL exists in Uranus since the time of its formation. Given the proximity of both values, however, further effects like planetesimal capture near the end of the formation process may have provided sufficient energy to overcome that energy threshold temporarily. Such a scenario would lend support to our assumption of an initially adiabatic planetary interior.

We argue that this boundary is stably stratified. Alternatively, one might think of a semiconvective transition zone. Semiconvection can take place if $1< R_{0}^{-1} < R_{\rm crit}$, $R_{\rm crit}:=(1+{\Pr}) / (\tau+{\rm Pr})$, where Pr is the Prandtl number, $\tau$ the ratio of particle to thermal diffusivities \citep{Rosenblum11},
and $R_0^{-1}:= (\alpha_{\mu}/\alpha_T)\nabla_{\mu}/ (\nabla_T - \nabla_{ad})$ so that $1<R_{0}^{-1}$ repeats Equation \ref{eq:ledoux}.
While precise values of Pr and $\tau$ are not known for the ice giants, preliminary estimates yield $\tau < 1$ and $\rm Pr > 1$ along the rather cold Uranus adiabat (M.~French, pers.~comm.~2013) so that $R_{\rm crit}\approx 2$. Thus, favorable conditions for semi-convection in the thermal boundary layer region appear confined to a very small interval in parameter space only, though this picture may require revision once Pr and $\tau$ values for warm, fluid ices are available.

We consider that small window unlikely to have a more than negligible influence on the heat and particle flux across the TBL.

We have seen that the stability criterion provides an unrealistically high upper limit for the temperature change across the layer boundary. However, the temperature at the inner edge of the layer boundary is of high importance, both for the determination of the energy reservoir of the deep interior and for the internal $T$--$P$--$\rho$ profile, thus for the composition determination. In the next Section we attempt to determine  $\Delta T$ by adjusting it to the observed luminosity.
This enables us to construct the first Uranus structure model that is consistent with both the gravity and luminosity observational constraints.

%%%%%%%%%%%%%%%
\begin{figure}[t]
\centering
\includegraphics[width=0.45\textwidth]{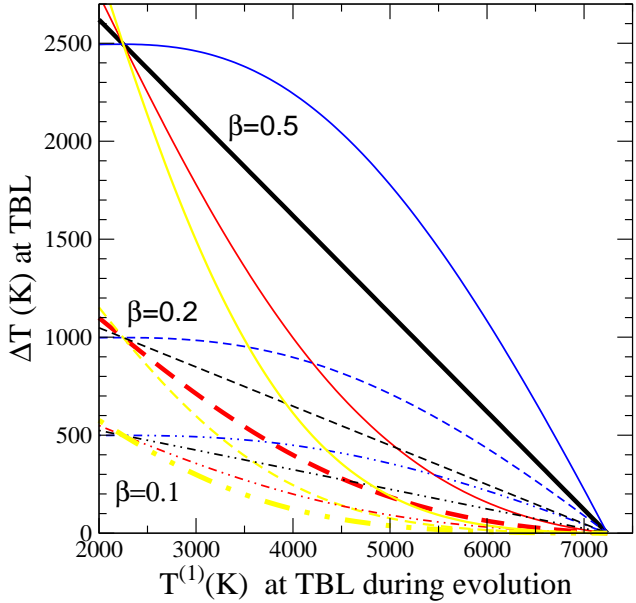}
\caption{\label{fig:deltaTfunctions}(Color online)
Assumed relations between $\Delta T$ and $T^{(1)}$ during the evolution of Uranus according to Equation \ref{eq:deltaT} with $X=X_{lin}$ (\emph{black}), $X=X_{red}$ (\emph{red}), $X=X_{blu}$ (\emph{blue}), and $X=X_{yel}$ (\emph{yellow}) at constant values of $\beta= 0.5$ (\emph{solid}), 0.2 (\emph{dashed}), and 0.1 (\emph{dot-dot-dashed}), see Equation \ref{eq:deltaT}. \emph{Thick lines} show relations that yield a cooling time $\tau\simeq\tau_{\odot}$.}
\end{figure}
%%%%%%%%%%%%%

%%% Hubbard et al 1995, magnetic field models %%%
\paragraph{Discussion} For comparison, \citet{Hubbard95} considered a thermal boundary layer which would cut off the heat flux from below a certain depth. Its location was estimated by matching the present luminosity, and was thus concluded to exist at much deeper levels of about 0.5 $\MU$, in disagreement with the structure models that match the gravity data. On the other hand, the result of  \citet{Hubbard95} was further used by \citet{StanBlox06} to compute thin shell magnetic dynamo models that indeed were able to reproduce the observed magnetic field well. This situation got even more complicated by the finding of convective thick-shell dynamo models that would as well reproduce the observed magnetic field \citep{Soderlund13}, while not necessarily being consistent with the low luminosity. Since many suggestions about the internal structure of Uranus exist that exhibit a wide range of consistencies and inconsistencies, we think a simplistic approach as pursued here is justified in order to learn what kind of structures could be consistent with all available constraints.

%%%%%%%%%%%%%%%%%%%%%%%%%%%%%%%%%%%%%%%%%%%%%%%%%%%%%%%%%%%%%%%%%%%%%%%%%%
\subsection{Evolution with adjusted thermal boundary layer}\label{sec:SIPevol}

%%%%%%%%%%%%%%%%%%%%%%%%%%%%%%%%%%%%%%%%
%{\bf \large[Figure \ref{fig:deltaTfunctions} (top left)]}
%\hspace*{1cm}\\
%%%%%%%%%%%%%%%%%%%%%%%%%%%%%%%%%%%%%%%

%%%%%%%%%%%%%%%%%%%%%%%%%%%%%%%%%%%%%%%%
%{\bf \large[Figure \ref{fig:cooltime_deltaT} (top right)]}
%\hspace*{1cm}\\
%%%%%%%%%%%%%%%%%%%%%%%%%%%%%%%%%%%%%%%

%%%%%%%%%%%%%%%
\begin{figure}[t]
\centering
\includegraphics[width=0.45\textwidth]{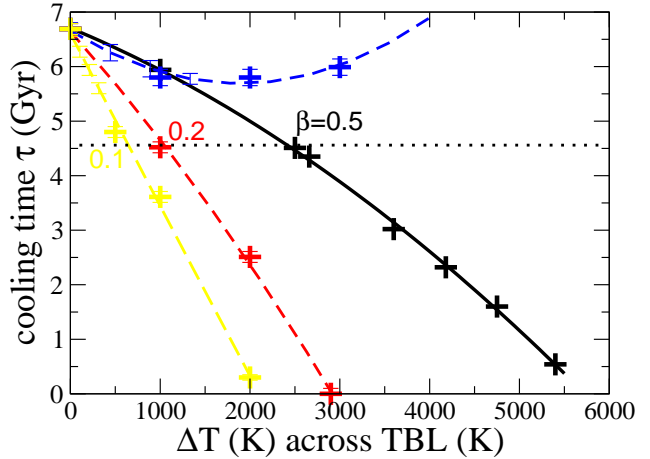}
\caption{\label{fig:cooltime_deltaT}(Color online) Cooling times of evolution models with thermal boundary layer (TBL)
against $\Delta T$ of present Uranus using different $\Delta T$--$T^{(1)}$ relations shown in Figure \ref{fig:deltaTfunctions}. Same colors code same relations.
Depending on the relation assumed, $\tau\simeq \tau_{\odot}$ is obtained for $\Delta T(t_0)$ ranging from 
$\approx 500$~K (\emph{yellow}) to 2500~K. The linear relation (\emph{black}) yields about the maximum value for 
$\Delta T(t_0)$.}
\end{figure}
%%%%%%%%%%%%%

We consider the change of temperature across the TBL, $\Delta T$, a free parameter and use it to adjust the luminosity of Uranus to the observed one at present time. For that purpose we introduce a scaling factor, $\beta$ and several relations between $\Delta T$ and $T^{(1)}$ that are supposed to approximately reflect the unknown behavior of a possible TBL in the real planet. In particular we assume 
\begin{equation}
\Delta T = \beta \times dT \times X\label{eq:deltaT}
\end{equation} with different options for $X$, $X_{lin}= 1$, $X_{red} = \exp[ -x ]$  
with $x=(T^{(1)}/T_{\rm m})^2 - (T^{(1)}(500) / T_{\rm m})^2$,
$X_{blu} = 2-\exp[ -x ]$, and $X_{yel} = \exp[ -2\, x]$, 
while 
\begin{equation}
dT = T^{(1)}(500\,K) - T^{(1)}(T_{\rm 1 bar})\:,   \label{eq:dT}
\end{equation}
with $T^{(i)}(T_{\rm 1 bar})$ being the temperature in layer No.~$i$ at the layer boundary, 
and $T_{m}=0.5\:T^{(1)}(76\, {\rm K}) + 0.5\:T^{(1)}(500\,{\rm K})$.
These relations are not meant to be exhaustive; rather, they serve us here as a proof of concept for Uranus.
Figure \ref{fig:deltaTfunctions} shows these relations.
The hottest profile considered is for $T_{\rm 1 bar}=500$ K, which we assume to be an adiabatic one with no change of temperature ($T^{(1)}=T^{(2)}$) but a change of composition at the mass coordinate $m_{12}$ of the layer boundary, which  
is kept at constant value as for homogeneous evolution models. 
While the outer, adiabatic and convective outer envelope cools efficiently over time, the deep interior below the TBL  does not. Therefore, the change of temperature grows over time, see Eq.~(\ref{eq:dT}). 
The higher the value of $\beta$, the less heat can escape. For $\beta=1$ no heat can escape from the deep interior below the layer boundary, while $\beta=0$ corresponds to the usual adiabatic case.
Thus we can use $\beta$ to adjust the computed luminosity of the thermal evolution model to the observed luminosity of Uranus, or, in other words, the computed cooling time to the known age of 4.56 Gyr. 

Figure \ref{fig:cooltime_deltaT} shows the cooling time obtained for different values of $\beta$ and the different assumed $\Delta T$--$T^{(1)}$ relations. 
For the linear case, we obtain $\tau_{\rm U}\approx\tau_{\odot}$ for $\beta\approx 0.5$. This yields $\Delta T=2500\:$K for present Uranus. We consider this a reasonable result since for comparison, the temperature change at the Earth core-mantle boundary is estimated to be 500--1000 K \citep{Poirier2000}.

The linear case appears to yield an upper limit to $\Delta T$. If, otherwise, $\Delta T$ would increase  more rapidly while the planet is still young (blue curve in Figure \ref{fig:deltaTfunctions}) deep internal heat cannot escape although the then warm atmosphere would help to do so.  At late times, if the same $T$--$P$ profile is to be reached as in the linear case, more heat from the large deep internal energy reservoir must be allowed to pass the TBL ($\beta < 1$) and go through the then cold atmosphere. This prolongs the cooling time compared to the linear case. Such a behavior was previously found for Saturn under the assumption of heat being retained in the deep interior by a semi-convective TBL \citep{LC13}.

Conversely, if $\Delta T$ would stay smaller than in the linear case at early times, much heat can be lost, as in the adiabatic case. This reduces the necessary reduction in heat loss at present time (red and yellow curves in Figure
\ref{fig:cooltime_deltaT}). The value of $\Delta T$ in the real planet will depend on the thermal conductivity. Diffusivity values of particles, the carriers of diffusive heat transport, under planet interior conditions are found to  increase with temperature \citep{Wilson15}, suggesting more efficient loss across the TBL at young than at old ages. We therefore disfavor the blue profile in Figure \ref{fig:deltaTfunctions} and consider the linear profile a proxy for the upper limit. We caution, however, that the real temperature profiles and thus the jump at the TBL can be different from those considered here once the assumption of an fully adiabatic deep interior is relaxed.

%%%%%%%%%%%%%%%
\begin{figure}
\centering
\includegraphics[width=0.45\textwidth]{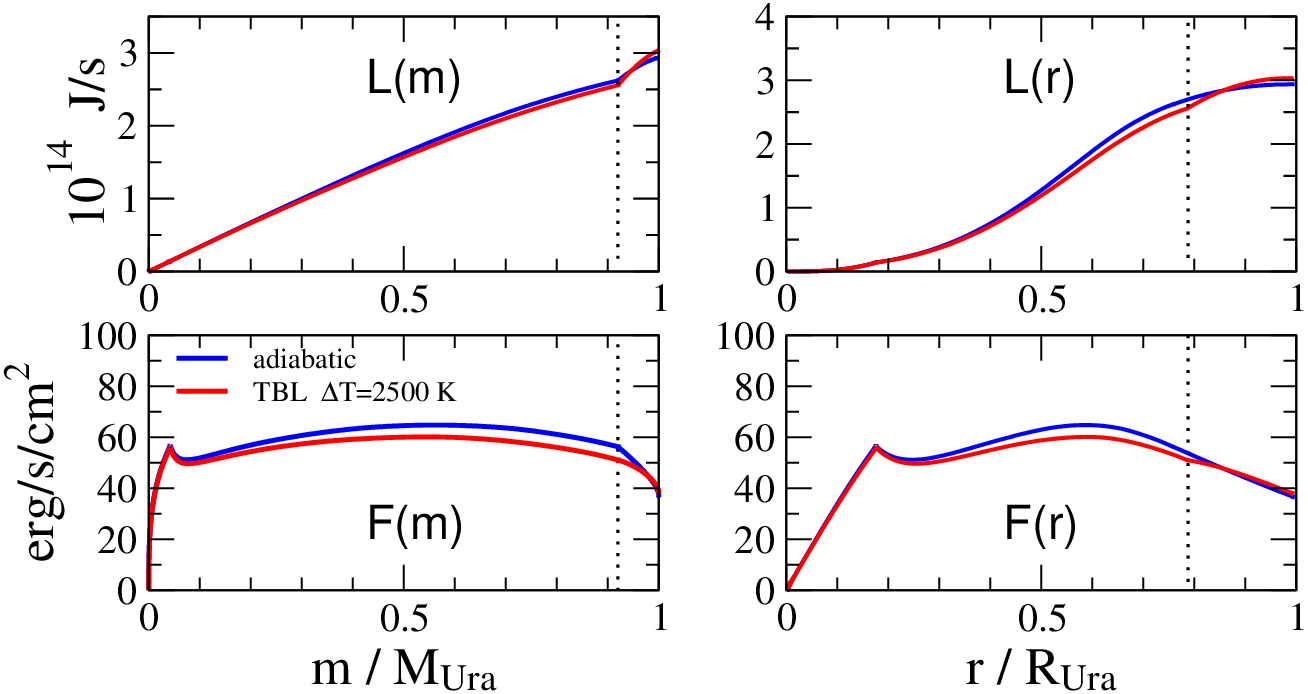}
\caption{\label{fig:lumiflux}Internal profiles of luminosity ($L$) and heat flux ($F$) for the class II thermal evolution
model with linear functional form for $\Delta T$ (\emph{red}) in comparison to an adiabatic standard model (\emph{blue}). The vertical dotted line denotes the location of the TBL.}
\end{figure}
%%%%%%%%%%%%%

%%%%%%%%%%%%%%%%%%%%%%%%%%%%%%%%%%%%%%%%
%\hspace*{1cm}\\
%{\bf \large[Figure \ref{fig:lumiflux}]}
%\hspace*{1cm}\\
%%%%%%%%%%%%%%%%%%%%%%%%%%%%%%%%%%%%%%%

\paragraph{Heat flux}
One might expect a big uncertainty in the heat flux  through the planet, $F(m)$ as a result of the significant uncertainty in the temperature profile around the TBL. We have computed heat flux profiles $F(m) = L(m)/4\pi r^2\rho(r(m))$
from the luminosity profiles $L(m)$ for both the adiabatic case and the model for the TBL according to the linear case.  The difference in the heat flux turns out to be small,  see Figure \ref{fig:lumiflux}. In fact, we find that the heat flux throughout Uranus' interior is close to the observed value in the atmosphere, $\Fint=45\pm 47$ erg/s/cm$^2$. From this investigation we conclude that the internal heat flux is well constrained by the assumed surface value, here $\Fint=40$ erg/s/cm$^2$ and that the temperature profile adjusts itself to be able to transport that heat.

%%%%%%%%%%%%%%%%%%%%%%%%%%%%%%%%%%%%%%%%%%%%%%%%%%%%%%%%%%%%%%%%%%%%%%%%%%%%%%%%%%%%%%%%%%%%%%%%%%%%%
\section{Evolution III: adjusted isothermal atmosphere}\label{sec:evol3}

%%%%%%%%%%%%%%
\begin{figure}
\centering
\includegraphics[width=0.45\textwidth]{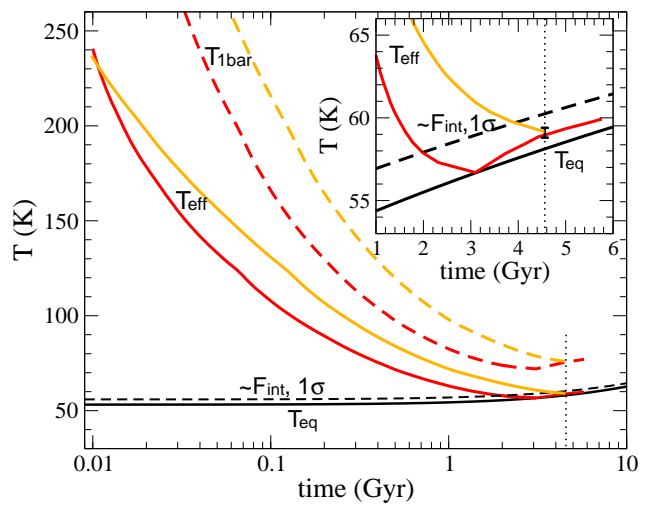}
\caption{\label{fig:evolU_teq}(Color online) Uranus cooling curves. The Class III model (\emph{red}) 
assumes that $\Teff$ begins to follow $\Teq$ (\emph{solid black}) prior to present time (\emph{vertical dotted}).  
It can be realized by furthermore assuming a TBL, and a subadiabatic zone in the deep atmosphere 
(Section \ref{sec:evol3}). 
For the Class II models (\emph{yellow}), the thermal boundary layer is adjusted to yield $\Teff\approx\Tint$ right at present time (Section \ref{sec:SIPevol}). Finally, the \emph{black dashed} line indicates the measured $1\sigma$ uncertainty of $\Fint$.}
\end{figure}
%%%%%%%%%%%%

%%%%%%%%%%%%%%%%%%%%%%%%%%%%%%%%%%%%%%%%
%{\bf \large[Figure \ref{fig:evolU_teq}]}
%\hspace*{1cm}\\
%%%%%%%%%%%%%%%%%%%%%%%%%%%%%%%%%%%%%%%%

In the previous Section, we have constructed an interior model that matches the observed luminosity at the present time.
That model implies that we live in a special epoch: the epoch when Uranus' effective temperature begins to follow its equilibrium temperature,  see the \emph{yellow curve} in \fig\ref{fig:evolU_teq}. In that scenario, the present time would separate the future, billions-of-years long near-equilibrium evolution ($L_{\rm eff} \simeq L_{\rm eq}$), which is also the thermal state of the Earth, from the billions-of-years long past of efficient cooling ($\Leff \gg \Leq$).
We estimate the probability to live in that epoch by $\Delta t_{\rm Teff}/\tau_{\odot}$ $\approx$ 0.5 Gyr/4.56 Gyr $=$ 11\%, 
where $\Delta t_{\rm Teff}$ is the duration spent at an $\Teff$ value within its observational $1\sigma$ uncertainty. Thus, if we would have observed Uranus already at time $t_0-\Delta t_{\rm Teff}$ we would have derived $\Teff$ and $\Tint$ values in agreement with the current respective  $1\sigma$ confidence intervals. While not really low, an 11\% probability would nevertheless distinguish the current epoch from earlier times. 

In this Section, we aim to find a solution that does not particularly highlight the current epoch, and thus has a higher probability of being what was the realized path. Therefore, we allow the era of $\Leff \gtrsim \Leq$ to begin at any time in the past. This property requires the presence of a TBL in order to obtain $\tau_{\rm U} \approx \tau_{\odot}$. In our representation of the TBL as described in Section \ref{sec:SIPevol}, this implies $\beta > 0$, and for the linear case in particular $0.5 < \beta < 1$. The higher the value of $\beta$, the earlier the start of the equilibrium evolution. For our purpose of illustrating an alternative scenario for Uranus, we chose $\beta=0.9$, which yields a cooling time of 3 Gyr until $\Teff\gtrsim\Teq$ sets in.

Figure \ref{fig:evolU_teq} shows such a model (\emph{red curves}). While $\Teff$ of the young and warm Uranus decreases with time, as it is the case for any isolated planet, its $\Teq$ increases due to the rising luminosity of the Sun over the course of its main-sequence evolution. As long as $\Teff > \Teq$, Uranus' outer envelope can cool efficiently.  From the time that $\Teff \simeq \Teq$ happens on, $\Teff$ follows $\Teq$ and thus rises with time. We represent this behavior by  setting $\Teff$ arbitrarily to a value which ensures that the intrinsic flux stays within the measured $1\sigma$ uncertainty, see \fig\ref{fig:evolU_teq}.

%%%%%%%%%%%%%%%%%%%%%%%%%%%%%%%%%%%%%%%%%%%%%%%%%
\subsection{Deep atmospheric temperature profile}

Once $\Teff$ starts to increase with time, so must the temperature in the atmospheric regions from which the intrinsic heat is radiated away. Typically, for Jovian planets these are regions of low optical depth at pressures below 1 bar \citep{Ingersoll87}. In particular, according to the here applied Graboske model atmosphere for that region, $\Teff$ is directly related to the 1-bar temperature. Thus, $T_{\rm 1 bar}$ must rise with time, too. 

%%%%%%%%%%%%%%
\begin{figure}
\centering
\includegraphics[width=0.45\textwidth]{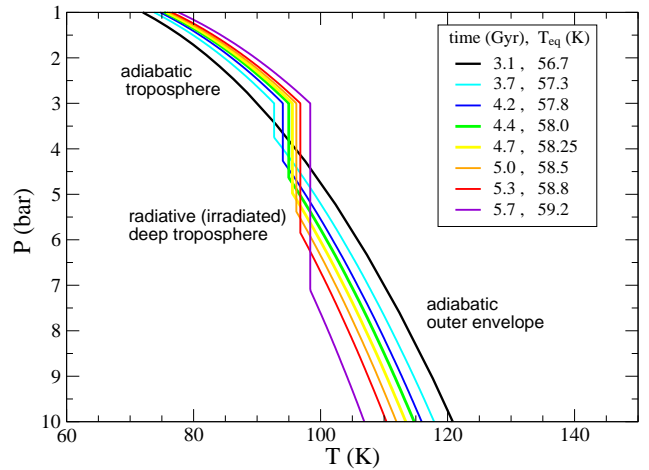}
\caption{\label{fig:atmPTevol}(Color online) Atmospheric $P$--$T$ profiles during Uranus' equilibrium evolution. 
A radiative zone, for simplicity assumed to be isothermal, extends between 3 bar and a level determined from its 
present heat flux.} 
\end{figure}
%%%%%%%%%%%%

%%%%%%%%%%%%%%%%%%%%%%%%%%%%%%%%%%%%%%%%%
%\hspace*{1cm}\\
%{\bf \large[Figure \ref{fig:atmPTevol}]}
%\hspace*{1cm}\\
%%%%%%%%%%%%%%%%%%%%%%%%%%%%%%%%%%%%%%%%

Contrary, the deeper interior will not warm up with time as that would require an additional heat source other than the Sun to conserve the total energy. As a result of rising surface temperatures  but slowly cooling deep interior, the atmosphere below the 1 bar level can no longer stay fully adiabatic but must develop a zone of shallower-than-adiabatic temperature profile, i.e.~a radiative zone. Eventually, any solar system giant planet will be gripped by such a fate. In case of Saturn for instance, \citet{Fortney07a} predict a subadiabatic, radiative zone in the atmosphere once Uranus-like low $\Tint$ values of 30 K or less are reached (See their Figure 2). There, it is seen to develop at about 30 bars.

For Uranus, it is unknown at what depths such a zone can reside. We let it arbitrarily start at 3 bar, emphasizing that there is nothing inherently special about this choice, rather than 30 bars or so.
In particular, our choice of 3 bar is not inconsistent with the $P$--$T$ profiles determined from atmosphere observations, because levels of 3 bars and below are difficult to probe observationally. For instance,  the temperature profile retrieved from the Voyager infrared spectral measurements can be described by an adiabatic profile down to about 1 bar \citep{GierCon87}, and the temperature profile derived from the Voyager radio occultation experiment indicates a steep, perhaps superadiabatic temperature gradient down to only 2.3 bar \citep{Lindal87}.

%%% figure 11: isothermal region during evolution %%%
For simplicity, we represent that radiative, subadiabatic zone by an isothermal temperature profile. 
The depth of the radiative zone is chosen to yield $\Teq \simeq \Teff$.
The resulting atmospheric $P$--$T$ profiles during the equilibrium evolution are shown in \fig\ref{fig:atmPTevol}. An interesting result is that the radiative region would be shallow, with bottom pressures of less than 10 bars today  if starting at 3 bars.

 %%%%%%%%%%%% origin of the low luminosity %%%%%%%%%%%%
In our model, the radiative zone in the atmosphere results from the combination of efficient cooling of the outer envelope in the past as a result of the internal structure, leading to low $\Teff$ at present, \emph{and} strong irradiation at present, i.e.~high $\Teq$, together yielding a low 
$\Tint^4=\Teff^4-\Teq^4\simeq 0$. This is similar to but not exactly as proposed by \cite{Hubbard78} who sees the origin of the low luminosity in a radiative zone in the atmosphere solely as a result of the strong incident flux. 

%%%%%%%%%%%%%%%%%%%%%%%%%%%%%%%%%%%%%%%%
%\hspace*{1cm}\\
%{\bf \large[Figure \ref{fig:profilePT}]}
%\hspace*{1cm}\\
%%%%%%%%%%%%%%%%%%%%%%%%%%%%%%%%%%%%%%%%

%%%%%%%%%%%%%%
\begin{figure}
\centering
\includegraphics[width=0.45\textwidth]{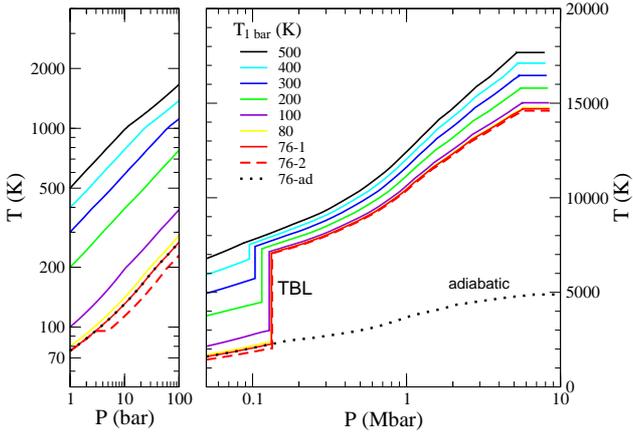}
\caption{\label{fig:profilePT}(Color online) Internal $P$-$T$ profiles of a Class III thermal evolution model. The underlying models have a superadiabatic TBL at $\approx 0.1$ Mbar (\emph{right panel}), 
and for the Uranus at present time a thin, subadiabatic radiative zone in the atmosphere at $\approx 5$ bars
(\emph{red dashed curve}). The black dotted curve shows an adiabatic profile.} 
\end{figure}
%%%%%%%%%%%%

Finally, in \fig\ref{fig:profilePT} we plot the $T$--$P$ profiles in Uranus' entire interior during its evolution according to the assumptions of this Section. With $\Tcore=15,000 K$, the central temperature might be up to 3 times higher than predicted by adiabatic models.

%%%%%%%%%%%%%%%%%%%%%%%%%%%%%%%%%%%%%%%%%%%%%%%%%%%%%%%%%%%%%%%%%%%%%%%%%%%%%%%%%%%%%%%%%%%%%%%
\section{Structure models with thermal boundary layer}\label{sec:structure}

We use the temperature profiles of our class II and III evolution models to construct standard three-layer structure models that match the gravity data using the method of \citet{N13UN}.

According to general EOS properties, %(except at anomalous phase transitions), 
warmer temperatures tend to decrease the density at given pressure level. Therefore, in order to  obtain a density profile that reproduces the gravity data, the composition of a warmer inner envelope must be different from the adiabatic case.  Higher densities can be achieved by a larger fraction of heavy elements, or by assuming elements of heavier atomic weight. As the inner envelope ice-mass fraction of our rock-poor adiabatic Uranus models with solar O:C:N ratios is already close to 100\%, it is necessary to allow for the presence of rocks in the inner envelope, while H/He looses its status as a necessary component in the deep interior. These two properties constitute important differences to the adiabatic models: these can be rock-free but must have some H/He in the deep interior \citep{Helled11,N13UN}. 

As with the adiabatic models, the lower limit of the I:R ratio can be zero according to the gravity data, in which case the deep interior would become a sole mixture of rocks and H/He.  
However, neither the mixing behavior of rocks with H/He nor of rocks with ices under Uranus interior conditions is well-understood. \citet{WiMi-MgO13} find miscibility of MgO with hydrogen at temperatures in excess of 10,000 K, which have probably occurred in the young Uranus. They may still occur today if a thermal boundary layer exists, but not if the entire planet is adiabatic, in which case central temperatures reach only 4500-6000 K. This property constitutes a third important  difference to the adiabatic models.  It supports the assumption that rocks can be homogeneously and linearly mixed with H/He in the deep interior of ice-giant planets. 

%%%%%%%%%%%%%%%%%%%%%%%%%%%%%%%%%%%%%%
%\hspace*{1cm}\\
%{\bf \large[Figure \ref{fig:tortenU}]}
%\hspace*{1cm}\\
%%%%%%%%%%%%%%%%%%%%%%%%%%%%%%%%%%%%%%
%%%%%%%%%%%%%%
\begin{figure}
\centering
\includegraphics[width=0.45\textwidth]{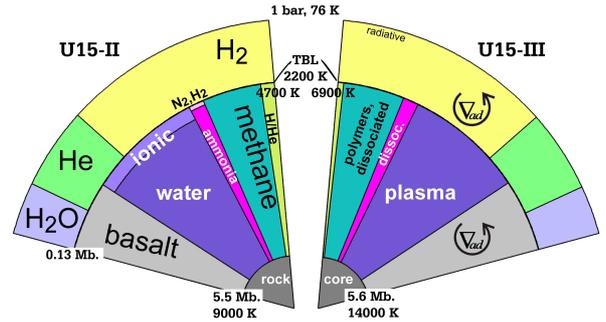}
\caption{\label{fig:tortenU}(Color online) Uranus three-layer structure models with thermal boundary layer that fit the gravity data and the luminosity. (\emph{Left}) Model U15-II, with a maximum change of temperature $\Delta T=2500$ K, see Section \ref{sec:SIPevol}; (\emph{right}) class III model U15-III, which has $\Delta T=4700$ K and a radiative atmosphere, see Section \ref{sec:evol3}. Colors indicate the composition by mass abundance as a function of radius, including hydrogen (yellow),  helium (green), water (blue) as a proxy for ices in the outer envelope,  and in the inner envelope H/He  (green-yellow), water (blue), ammonia (pink), methane (cyan), and rocks (gray). Overplotted are the single component phases of water (molecular, ionic, plasma), ammonia (molecular N$_2$,H$_2$, dissociated), and methane (polymeric/dissociated).}
\end{figure}
%%%%%%%%%%%%%

Our favored structure models are illustrated in Fig.~\ref{fig:tortenU}. They have $1\times$ solar I:R ratios and a 
small amount of H/He in the inner envelope. This is consistent with core accretion formation models, which predict the simultaneous accretion of small amounts of gaseous H/He in addition to the ice/rock planetesimals onto the protocore \citep{Pollack96,Mordasini12}.

%%% phase diagrams %%%
The maximum change of temperature at the layer boundary at 0.1 Mbar leads to $\approx 5000\,$ K ($\approx 9000\,$ K) higher central temperatures for the class II (class III) models compared to our adiabatic models. Regions where ionic water could occur in adiabatic models here reduce to a very thin shell or entirely disappear at the favor of the plasma phase \citep{Redmer11}. This may have implications on the viscosity and the electrical conductivity, and thus for the magnetic field generation \citep{Chau11,Soderlund13}. Moreover, pure ammonia would be in the dissociated phase \citep{Bethkenhagen13}, where it is also seen to be miscible with water in 1:1 ammonia-water mixtures \citep{Bethkenhagen15}. If the change of temperature is not too strong, molecular N$_2$ and O$_2$ could occur near the inner edge of the boundary layer. According to the experimentally determined methane phase diagram of \citet{Hirai09}, methane would undergo several states of polymerization in the inner envelope of Uranus, and possibly dissociate into hydrogen and diamond. The phase diagram and the thermophysical properties (diffusivity, viscosity, conductivity) of the full mixture are topics of current investigations \citep{Gao10,Chau11}. Our Uranus models can help to outline the regions of interest to be explored. We suggest exploration at 0.1--6 Mbar and 1000--15000 K. This is required for the further development of consistent models of the thermal evolution, formation, internal structure, and magnetic field generation of Uranus and Neptune.

%%%%%%%%%%%%%%%%%%%%%%%%%%%%%%%%%%%%
\section{Summary}\label{sec:summary}

In this paper we have considered the question of why Uranus is so faint. This observed property contains important information on  the internal structure, composition, and formation of Ice Giant Planets. 
To address this issue, we  have used ab initio EOS data for water, ammonia, and methane and computed three classes of Uranus evolution models.

First, we showed that higher rock mass fractions of conventional adiabatic models lead to further prolongation of the cooling time far beyond the age of the solar system. This is in opposition to what previous work had shown \citep{HMacF80}.

Second, we argued that the transition region between the  H/He rich outer envelope and the icy/rock rich deep interior, as indicated by the gravity data, should be stably stratified and thus act as a thermal boundary layer to the deep internal heat flux. Therefore, we constructed evolution models with a simple TBL (class II) and found that Uranus' low luminosity can be explained if the change of temperature across it is drastic, up to about $\Delta T\approx 2500$ K.
Such evolution models assume --like  the conventional models do-- that the current epoch separates a brighter past from a faint future. 

Third, we investigated the scenario that Uranus cooled down to its equilibrium temperature ($\Teff\simeq \Teq$) at some time in the past. For this class of models we find that Uranus should not only have a thermal, 
\emph{superadiabatic} boundary but also a shallow, radiative, \emph{subadiabatic} zone in the irradiated atmosphere at a few bars or deeper. 

%%% structure models
Our class II and III models yield by  a factor of up to about 2 to 3 warmer core temperatures than the class I models. As a result, the presence of rocks is required in the inner mantle in order to match the gravity data. We have shown that in both cases, structure models with assumed solar I:R ratios have non-zero deep internal H/He abundances, in agreement with core accretion planet formation models.

\paragraph{Outlook}
Our presented simple structure and evolution models leave plenty of room for improvement.
[1] The temperature and composition profiles should be calculated by considering the diffusive fluxes of heat and of the single particle species during evolution, as well as the possible flux enhancements at the interfaces between convective and diffusive regions. 
[2] The resulting possible locations of stable regions should be compared against magnetic dynamo models.
[3] If the TBL is not sharp as modeled here but extended far outward, our class III models may rapidly apply. Different widths should thus be studied.
[4] As the fluxes depend on the thermal conductivity and on the diffusion coefficients, we suggest to determine these transport properties for H, He, O, N, and C in mixtures of ices. 
[5] Strong constraints on the internal structure and evolution are expected to come from a better understanding of 
the miscibility behavior of rocks (MgO, FeO, SiO$_2$) with H/He and ices, as well as of carbon in an HNO-rich environment. Sedimentation of little carbon clusters \citep{Chau11} and rising of H-rich material may further affect the location of stable and of electrically conducting regions.
[6] The model atmosphere could be expanded to include the effect of equilibrium evolution, so that the depth of the radiative zone in Uranus' atmosphere can be determined and the in Section \ref{sec:evol3} proposed scenario  be evaluated. 
Furthermore, 
[7] Clouds probably play an essential role in the heat that can escape the planet, a phenomenon which is easily experienced on Earth. Thus, studying the effect of deep-seated water clouds \citep{Wiktorowicz07} on the thermal evolution may shed new insight on the origin of the faintness.
[8] Finally, any planetary modeling scheme for Uranus should be checked against Neptune.

At the current stage, the reason for the different heat fluxes of Uranus and Neptune remains an open question. 
Models for their internal structure suggest similar interiors. Therefore, also the luminosity of Neptune could be influenced by the presence of a thermal boundary layer, which in principle can both shorten (Uranus case; this work), or prolong (Saturn case; \citealp{LC13}) the cooling time, although current evolution models do not require one to explain Neptunes' luminosity \citep{Fortney11}.
Working through the outlined steps [1--7] can help to better understand these two ice giants and to develop useful models for planets of similar mass and size. Given recent suggestions for new missions to Uranus and Neptune \citep{Arridge11,Masters13} we think this is a good time to revisit important but still unexplained spacecraft based observational data such as the luminosity, as well as the interior models and their input physics that are supposed to explain them.

\subsubsection*{Acknowledgments}

We thank the two anonymous referees for constructive comments. NN thanks R.~Helled and M.~Podolak for interesting conversations, and participants of the Workshop on Ice Giant Planets 2014 in Laurel, MD, for fruitful discussions. We gratefully acknowledge the funding support from NASA under Contract No.~NNH12AU44I. MB and RR acknowledge support from the German Science Foundation (DFG) via SFB 652 and the computation time provided by the North-German Supercomputing Alliance (HLRN) and the ITMZ of the University of Rostock. . 
JJF acknowledges support from NSF grant AST-1010017 and NASA grant NNX11AJ40G-001.

%%%%%%%%%%%%%%%%%%%%%%%%%%%%%%%%
\bibliographystyle{model2-names}
\bibliography{msUN2-refs}

%%% Table 1 %%%%%%%%%%%
%\input{tab_layerdefs}
%%%%%%%%%%%%%%%%%%%%%%%

%%% Table 2 %%%%%%%%%%%
%\input{tabSIP_Ledoux}
%%%%%%%%%%%%%%%%%%%%%%%

%%%%%%%%%%%%

\end{document}